\documentclass[
hidelinks,
reprint,
superscriptaddress,
altaffilletter,
nofootinbib,
amsmath,
amssymb,
aps,
prd,
]{revtex4-1}

\usepackage{times}
\usepackage{mathtools}
\usepackage{orcidlink}
\usepackage{graphicx}
\usepackage{dcolumn}
\newcolumntype{d}[1]{D{.}{.}{#1}}
\usepackage{bm}
\usepackage{verbatim} 
\usepackage{cprotect} 
\usepackage{url}
\usepackage{subfigure} 
\usepackage{tikz}
\usepackage{afterpage}
\usepackage{nicefrac}
\usepackage{tabularx}
\newcolumntype{L}{>{\raggedright\arraybackslash}X}
\usepackage{multirow}
\usepackage[english]{babel}
\usepackage{natbib} 
\usepackage{glossaries} 
\usepackage{hyperref}
\hypersetup{
    colorlinks,
    linkcolor={blue!60!black},
    citecolor={blue!60!black},
    urlcolor={blue!60!black}
}
\usepackage{aas_macros}

\graphicspath{{figures/}}

\newcommand{\DocumentID}{P1900059}
\newcommand{\gstlal}{\texttt{GstLAL}} 
\usepackage{xcolor} 
\begin{document}

\title{Targeted subthreshold search for strongly lensed gravitational-wave events}
\author{Alvin K. Y. Li\,\orcidlink{0000-0001-6728-6523}}
\email{kli7@caltech.edu}
\affiliation{Department of Physics, The Chinese University of Hong Kong, Shatin, New Territories, Hong Kong}
\affiliation{LIGO Laboratory, California Institute of Technology, Pasadena, CA 91125, USA}

\author{Rico K. L. Lo\,\orcidlink{0000-0003-1561-6716}}
\email{kllo@caltech.edu}
\affiliation{LIGO Laboratory, California Institute of Technology, Pasadena, CA 91125, USA}

\author{Surabhi Sachdev\,\orcidlink{0000-0002-0525-2317}}
\affiliation{LIGO Laboratory, California Institute of Technology, Pasadena, CA 91125, USA}
\affiliation{Department of Physics, The Pennsylvania State University, University Park, PA 16802, USA}
\affiliation{Institute for Gravitation and the Cosmos, The Pennsylvania State University, University Park, PA 16802, USA}
\affiliation{Center for Gravitation and Cosmology, Department of Physics, University of Wisconsin, Milwaukee, P.O. Box 413, Wisconsin, 53201, USA}

\author{Juno\,C.L.Chan}
\affiliation{Department of Physics, The Chinese University of Hong Kong, Shatin, New Territories, Hong Kong}
\affiliation{Niels Bohr International Academy, Niels Bohr Institute, Blegdamsvej 17, 2100 Copenhagen , Denmark}

\author{E. T. Lin\,\orcidlink{0000-0002-0030-8051}}
\affiliation{Institute of Astronomy, National Tsing Hua University, Hsinchu 30013, Taiwan}

\author{Tjonnie G. F. Li\,\orcidlink{0000-0003-4297-7365}}
\affiliation{Department of Physics, The Chinese University of Hong Kong, Shatin, New Territories, Hong Kong}
\affiliation{Department of Electrical Engineering (ESAT), KU Leuven, Kasteelpark Arenberg 10, B-3001 Leuven, Belgium}

\author{Alan J. Weinstein\,\orcidlink{0000-0002-0928-6784}}
\affiliation{LIGO Laboratory, California Institute of Technology, Pasadena, CA 91125, USA}

\date{Received 9 December 2019; revised 9 December 2022; accepted 9 February 2023; published 13 June 2023}

\begin{abstract}
	Strong gravitational lensing of gravitational waves can produce duplicate signals separated in time with different amplitudes. We consider the case in which strong lensing produces superthreshold gravitational-wave events and weaker subthreshold signals buried in the noise background. We present the GstLAL-based TargetEd Subthreshold Lensing seArch search method for the subthreshold signals using reduced template banks targeting specific confirmed gravitational-wave events. We perform a simulation campaign to assess the performance of the proposed search method. We show that it can effectively uprank potential subthreshold lensed counterparts to the target gravitational-wave event. We also compare its performance to other alternative solutions to the posed problem and demonstrate that our proposed method outperforms the other solutions. The method described in this paper has already been deployed in the recent LVK Collaboration-wide search for lensing signatures of gravitational waves in the first half of LIGO/Virgo third observing run O3a [R. Abbott et al. (LIGO Scientific, Virgo Collaborations), \href{https://doi.org/10.3847/1538-4357/ac23db}{Astrophys. J. 923, 14 (2021)}.].
\end{abstract}

\maketitle

\section{Introduction}\label{Section: Introduction}
General relativity predicts that waves emitted from a source can be deflected before reaching the observer, an effect known as gravitational lensing~\cite{Bartelmann:2010fz}, which has been extensively studied with electromagnetic waves~\cite{Rahvar:2018nhx,Cunha:2018acu,Turyshev:2017pgm,Dokuchaev:2018fze,Bartelmann:2016dvf,Narayan:1996ba,Metcalf:2018elz,Clowe:2003tk,Markevitch:2003at,Bond:2004qd,Coe:2012wj}. Since the first successful detection of gravitational waves by the LVK Collaboration~\cite{LIGOScientific:2018mvr,LIGOScientific:2020ibl,LIGOScientific:2021usb,LIGOScientific:2016vbw,LIGOScientific:2016vlm,LIGOScientific:2017vwq,LIGOScientific:2020aai,LIGOScientific:2020ufj,LIGOScientific:2020zkf,LIGOScientific:2020stg,LIGOScientific:2021qlt}, works have started considering gravitational-wave lensing, including lensing rates~\cite{Ng:2017yiu,Li:2018prc,Oguri:2018muv,Buscicchio:2020cij,Mukherjee:2020tvr,Buscicchio:2020bdq}, strong-lensing effects~\cite{Wang:1996as,Dai:2017huk,Ezquiaga:2020gdt,Ng:2017yiu,Li:2018prc,Oguri:2018muv,Smith:2017mqu,Smith:2018gle,Smith:2019dis,Robertson:2020mfh,Ryczanowski:2020mlt}, and weak and microlensing effects~\cite{Deguchi:1986zz,Nakamura:1997sw,Takahashi:2003ix,Cao:2014oaa,Jung:2017flg,Lai:2018rto,Christian:2018vsi,Dai:2018enj,Diego:2019lcd,Diego:2019rzc,Pagano:2020rwj,Cheung:2020okf,Mishra:2021abc} on gravitational waves. Claims or disclamations of detected pairs of gravitational waves for being lensed images have also been made~\cite{Broadhurst,Singer:2019vjs,Diego:2021fyd}.

The LVK Collaboration recently published its first full- scale analysis to search for gravitational-lensing signatures of gravitational waves within data from the first half of LIGO/Virgo third observing run O3a~\cite{LIGOScientific:2021izm}. They conclude that no compelling evidence was found for gravitational lensing to take place within O3a. In the paper, they consider the possibility that strong lensing produces multiple gravitational waves from the same sources. In one scenario some images are magnified and hence become identifiable as detections, and the rest are demagnified and thus are buried within the noise background. Through two independent search methods, they search for the latter subthreshold lensed counterparts to confirmed gravitational-wave detections by effectively reducing the noise background while keeping the targeted foreground constant. This paper explain in details one of the methods being used, namely the \gstlal-based TargetEd Subthreshold Lensing seArch (TESLA) pipeline. We provide an assessment to its performance in searching for potential subthreshold lensed counterparts to superthreshold gravitational waves.

The paper structured as follows: In Sec. \ref{Section: Background}, we provide a brief overview of how matched-filtering search pipelines work to search for possible gravitational-wave candidates, using \gstlal$\,$ as an example, as well as explaining the basics of gravitational lensing and the motivation to search for subthreshold strongly lensed gravitational waves. In Sec. \ref{Section: Methods}, we pose the problem of searching for potential subthreshold lensed counterparts, and introduce the TESLA pipeline's working principle. In Sec. \ref{Section: MDC}, we provide details of a mock data challenge performed to assess the performance of the TESLA pipeline, and compare its effectiveness to alternative proposals in solving the problem posed in the previous section. Finally, Sec. \ref{Section: Conclusion} summarizes the findings and discusses possible future work to improve the search sensitivity of the TESLA pipeline.

\section{Background}\label{Section: Background}
\subsection{Search for gravitational waves with matched-filtering based compact binary coalescences search pipeline: Using \gstlal$\,$as an example}
Searches for gravitational waves from compact binary coalescences (CBC) typically utilize matched-filtering search pipelines, including \gstlal~\cite{2017PhRvD..95d2001M,Sachdev:2019vvd,Hanna:2019ezx,2021SoftX..1400680C,Godwin:2020weu,Chan:2020fip,2021PhRvD.103h4047M,2020arXiv201102457M}, PyCBC~\cite{Allen:2005fk,Allen:2004gu,DalCanton:2014hxh,Usman:2015kfa,Nitz:2017svb}, MBTA~\cite{Adams:2015ulm,Aubin:2020goo}, and SPIIR~\cite{76e97cf9801544919973534ed7028b6a}. To better explain the rest of the paper, we will give a brief overview of such pipelines. In particular, we are using \gstlal$\,$ as the example since it is mainly used in this work.

\subsubsection{Populating the candidate event basis by matched filtering}
Waveforms of gravitational waves from compact binary coalescences are well-modeled. Specifically, the time evolution of a CBC waveform is empirically governed by intrinsic parameters of the source (e.g., the source component masses $m_1, m_2$ and dimensionless spins $\vec\chi_1, \vec\chi_2$). Denote the data stream $d(t)$ in the time domain as $d(t) = n(t) + h(t)$, where $n(t)$ represents noise and $h(t)$ represents a signal in the data (if it exists). We detect gravitational-wave signals by cross-correlating noisy data using template with known parameters. The cross-correlation [quantified by the signal-to-noise ratio (SNR)] timeseries\footnote{The SNR is evaluated at different times of the data, and the results are recombined to form a continuous time series.} for a given data stream with a specific waveform template $h^\text{template}_{i}(t)$ as~\cite{Allen:2005fk,2017PhRvD..95d2001M,Sachdev:2019vvd}
\begin{equation}
	x_i(t) = 2\int_{-\infty}^{+\infty} \hat{h}^\text{template}_{i}(\tau) \hat{d}(\tau + t) d\tau,
\end{equation} 
where the ``hat" above the template and the data means that they are whitened with the single-sided power spectral density (PSD) $S_n(f)$ in the frequency domain (denoted by a ``tilde") according to
\begin{equation}
	\hat{h}^\text{template}_{i}(t) = \int_{-\infty}^{+\infty} \frac{\tilde{h}^\text{template}_{i}(f)}{\sqrt{S_n(|f|)/2}} e^{2\pi i f t} df.
\end{equation}
If a signal $h$ is truly present in the data, the SNR will be maximized if it is cross-correlated with a template waveform that has precisely the same parameters as it does, and when they are perfectly aligned in time. We denote that as the optimal SNR $\rho_\text{opt}$, defined mathematically as
\begin{equation}
	\rho_\text{opt}^2 = \max_t\left[\hat{h}(t)^2\right].
\end{equation}
Note that the strains of CBC gravitational waveforms are inversely proportional to the effective distance $D_\text{eff}$ to the source, i.e.
\begin{equation}
	\tilde{h}(f) \propto \frac{1}{D_\text{eff}},
\end{equation}
with
\begin{equation}
	D_\text{eff} = D\left[F_+^2 \left(\frac{1+\cos^2 \iota}{2}\right)^2 + F_\times^2 \cos^2 \iota  \right]^{-1/2},
\end{equation}
where $F_+$ and $F_\times$ are the antenna response functions corresponding to the signal, $\iota$ is the inclination of the source relative to the line of sight, and $D$ is the luminosity distance to the source. The optimal SNR for a given template therefore scales inversely with the source's effective distance, i.e.
\begin{equation}\label{eqn:rho_distance}
	\rho_\text{opt} \propto \frac{1}{D_\text{eff}}.
\end{equation}
We will exploit such scaling in later parts of this work.

A large template bank containing a set of gravitational waveforms is used to cover the desired search parameter space in a general search. Within the search space, templates are not distributed uniformly, but instead in a way that satisfies a minimal match criterion to balance between identifying signals with minimal loss of SNR, and accumulating too much noise background. The general search space is wide since we have no prior information regarding the signal?s parameter subspace. For instance, the template bank used to search for gravitational waves in data collected by LIGO/Virgo detectors within the first half of the third observing run O3a consists of $1\mathrm{,}412\mathrm{,}263$ templates. The templates have component masses ranging from $1M_\odot$ to $400M_\odot$, covering signals from binary neutron stars, binary black holes and neutron-star black hole mergers~\cite{LIGOScientific:2020ibl,LIGOScientific:2021usb} (See Figures~\ref{Fig: MGW220111a targeted bank}, \ref{Fig: MGW220111a random bank} or \ref{Fig: MGW220111a PE bank}).

Prior to performing matched-filtering, \gstlal$\,$ further divides the template bank by grouping templates that will respond to noise in similar ways into sub-banks~\cite{2017PhRvD..95d2001M,Sachdev:2019vvd,Sachdev:2020lfd}. It then utilizes the LLOID method~\cite{Cannon:2011vi,Sachdev:2020lfd} to create orthogonal basis filters from the sub-banks through in-order multibanding and singular value decomposition (SVD)~\cite{PhysRevD.82.044025} for each of the time slices\footnote{Time slices are disjointly supported intervals in time within a template}. The basis filters are then are then used to perform matched-filtering through the data stream for each detector. The results are combined to reconstruct the SNR timeseries for each template. The SNR timeseries are then maximized over short time windows in order to produce a set of triggers\footnote{A trigger refers to a certain time in the data stream which gives an SNR larger than a threshold value.} for each template and each detector. To reduce the number of triggers, only those with an SNR greater than $4$ are kept to form the candidate event basis.

\subsubsection{Assigning statistical significances for the candidates}
In order to rank the candidates, \gstlal$\,$ assigns each of them a log likelihood-ratio $\ln \mathcal{L}$, defined by
\begin{equation}
	\ln \mathcal{L} = \ln \frac{P(\vec{D}_H, \vec{O}, \vec{\rho}, \vec{\xi}^2, [\Delta \vec{t}, \Delta \vec{\phi}]|\vec{\theta}, \text{signal})}{P(\vec{D}_H, \vec{O}, \vec{\rho}, \vec{\xi}^2, [\Delta \vec{t}, \Delta \vec{\phi}]|\vec{\theta}, \text{noise})}\cdot \frac{P(\vec{\theta}|\text{signal})}{P(\vec{\theta}|\text{noise})},
\end{equation}
which is the log of the probability ratio of obtaining the candidate event under the signal model versus the noise model. The $\ln \mathcal{L}$ ranking statistics depend on (1) the participating detectors $\vec{O}$, (2) the horizon distances (and hence the sensitivities) of the participating detectors $\vec{D}_H$, (3) the matched-filter SNRs $\vec{\rho}$ and (4) the auto-correlation based signal consistency test values of the event at each detector $\vec{\xi}^2$. For coincident events\footnote{That is, triggers that are found in multiple detectors within a certain time window.}, $\ln \mathcal{L}$ also depends on (5) the time delays $\Delta \vec{t}$ and (6) the phase delays $\Delta \vec{\phi}$ of the trigger between participating detectors, which are enclosed in square brackets in the above equation~\cite{2015arXiv150404632C,2017PhRvD..95d2001M,Sachdev:2019vvd,Hanna:2019ezx}. Starting from the analysis for O3a data, \gstlal$\,$ also includes a template-dependent factor, $P(\vec{\theta}|\text{signal})$ with $\vec{\theta}$ representing the template parameters, that reflects how consistent the trigger template parameters are with an assumed astrophysical mass model~\cite{Heather}.
\gstlal$\,$estimates the $\ln \mathcal{L}$ distribution for noise triggers by sampling the noise distributions of the parameters it depends on using Monte Carlo methods~\cite{LIGOScientific:2018mvr,2017PhRvD..95d2001M,Sachdev:2019vvd}.
After assigning the $\ln \mathcal{L}$ ranking statistics, \gstlal$\,$then evaluates, for each event, a false-alarm-rate (FAR) that quantifies how often noise can produce a trigger with a ranking statistic $\ln \mathcal{L}$ greater or equal to the ranking statistic $\ln \mathcal{L}*$ of the trigger under consideration, marginalized over all the data analyzed~\cite{LIGOScientific:2018mvr}. Mathematically, we have
\begin{equation}
	\text{FAR} = \frac{N\times \text{FAP}}{T},
\end{equation}
where $N$ is the total number of observed candidates, $T$ is the duration of the data being analyzed, and FAP, or false-alarm-probability, is the probability for which noise can produce a trigger with a ranking statistic $\ln \mathcal{L}$ greater or equal to the ranking statistic $\ln \mathcal{L}*$ of the trigger under consideration, defined mathematically as
\begin{equation}
	P(\ln \mathcal{L}>\ln \mathcal{L}^*|\text{noise}) = \int_{\ln \mathcal{L}^*}^{\infty} P(\ln \mathcal{L} | \text{noise}) d\ln \mathcal{L}. 
\end{equation}

\subsubsection{Outputting a list of candidate events for further analysis}
Finally, \gstlal$\,$produces a list of candidate events ranked by their evaluated ranking statistics for further analysis. The FARs assigned to each candidate event by \gstlal$\,$ quantifies how often noise fluctuations could generate the event under consideration. The lower the FAR, the more likely the event is a gravitational wave. It is up to the analysts to decide a threshold\footnote{That is to say, there does not exist a decisive cut in FAR that distinguishes gravitational-wave event triggers from noise triggers.} above which they would perform further analysis for a candidate event. In this work, we define superthreshold triggers as those with FAR $< 1/30$ days, while subthreshold triggers are required to have SNR$>4$. However, note that the FAR assignment also depends on the number of noise triggers found during the search, which depends on the number (and distribution) of templates used for the search. Increasing the number of templates to target a broader search space allows us to look for gravitational waves coming from a broader source population. However, this will also lead to a higher trials factor and hence larger noise background. Consequently, some potential (weaker) gravitational waves will have lower ranking statistics that might not pass the usual conservative threshold, and thus remain unidentified. Nevertheless, reducing the number of templates does not necessarily improve the ranking statistics for all potential gravitational waves, since it also depends on the template distribution, i.e., the search space that we are interested in.

In later sections of this paper, we target a smaller region of parameter space to search for potential weaker gravitational-wave signals within the data that could be lensed counterparts to a target superthreshold gravitational wave for further analysis.

\subsubsection{Parameter estimation: Determining the source parameters in a more refined manner}
While \gstlal$\,\,$and other aforementioned CBC search pipelines provide a list of candidate events with the source parameters of the accompanying templates that identify them, they should not be misunderstood to be providing a concrete estimation for the source parameters of each candidate event, since the sole purpose of the search pipelines is simply to identify possible gravitational-wave candidates. In order to obtain a more rigorous estimation for the source parameters, Bayesian parameter estimation (PE) is required. Details about how PE is done are out of the scope of this paper, and hence it will not be discussed extensively; interested readers should refer to \cite{Ashton:2018jfp,Romero-Shaw:2020owr,lalinference_o2,Veitch:2014wba}.It suffices to say here that PE outputs a set of posterior samples that provides the posterior probability distribution, which gives the best estimates of the source parameters for each candidate event analysed. In later sections of this paper, we will use the posterior samples for confirmed gravitational- wave events to reduce the search parameter space to look for potentially weaker gravitational-wave signals.

\subsection{Basics of strong lensing of gravitational waves}
Gravitational lensing refers to the effect predicted by general relativity that waves emitted from a source can be deflected due to the distortion of spacetime by the gravitational potential wells of massive objects (e.g., galaxies or galaxy clusters) before reaching the observer. Such effect has been long observed and investigated for electromagnetic (EM) waves~\cite{Rahvar:2018nhx,Cunha:2018acu,Turyshev:2017pgm,Dokuchaev:2018fze,Bartelmann:2016dvf,Narayan:1996ba,Metcalf:2018elz,Clowe:2003tk,Markevitch:2003at,Bond:2004qd,Coe:2012wj}. However, gravitational waves are no different from EM waves according to the equivalence principle, and hence should also be affected similarly by gravitational lenses. For the rest of this paper, we focus on strong lensing of gravitational waves assuming geometrical optics. That is, we assume the wavelength of gravitational waves is much shorter than the spatial extent of the potential well of the gravitational lenses, allowing one to neglect diffraction effects. Under such an assumption, strong lensing can produce repeated signals for transient gravitational wave coming from the same source separated by a relative arrival time delay $\Delta t_j$\footnote{For instance, galaxy lenses can produce repeated gravitational-wave signals coming from the same source separated by a time delay ranging from minutes to months \cite{Ng:2017yiu,Li:2018prc,Oguri:2018muv}.} with basically identical waveforms, apart from an overall scaling factor $\sqrt{\mu_j}$, that amplifies / deamplifies the signals, and an additional Morse phase factor, depending on the lensed signal type~\cite{Wang:1996as,Dai:2017huk,Ezquiaga:2020gdt,LIGOScientific:2021izm}. Mathematically, suppose $\tilde{h}(f;\vec{\theta}, \Delta t_j = 0)$ denotes the not-lensed gravitational waveform in the frequency domain with source parameters $\vec{\theta}$ (including the coalescence time $t_c$), the $j^\text{th}$ strongly-lensed counterparts will have waveforms $\tilde{h}^{L}_j$ given by
\begin{equation}
	\tilde{h}^{L}_j(f; \vec{\theta}, \mu_j, \Delta t_j, \Delta \phi_j) = \sqrt{|\mu_j|} \tilde{h}(f;\vec{\theta}, \Delta t_j) e^{(i \text{sign}(f) \Delta \phi_j)}, 
\end{equation} 
where $\Delta t_j$ denotes the time delay relative to the not-lensed signal's coalescence time, $\sqrt{\mu_j}$ is the amplitude scaling factor due to lensing magnification / demagnification, and $\Delta \phi_j$ is the additional Morse phase factor, given by
\begin{equation}
	\Delta \phi_j = -\frac{n_j \pi}{2},
\end{equation}
with $n_j=0, 1$ and $2$ for Type I, II and III lensed signals, corresponding to a minimum point, saddle point and maximum time-delay solution to the lens equation respectively. Note that the magnification factor $\sqrt{\mu_j}$ (1) is frequency-independent under the assumption of geometric optics, and (2) can take on values larger or smaller than $1$, i.e. the lensed signals can become either stronger or weaker in amplitude as compared to the not-lensed waveform.
Note that while images produced from strong lensing will appear to be at different sky locations, the difference (in order of arc-seconds) is negligible compared to the uncertainty in sky localization for gravitational waves (in order of degrees). Hence, throughout this work we assume multiple gravitational-wave images from the same source will appear to come from essentially the same sky location. 
To summarize, strongly-lensed gravitational waveforms are identical (with the same intrinsic parameters, i.e. masses and spins, and sky location) to the not-lensed one apart from (1) a relative arrival time delay, (2) an overall scaling factor which can either magnify or de-magnify the signal, and (3) an additional Morse phase factor.

\subsection{Search for subthreshold lensed gravitational-wave signals}
The lensing magnification factor can take on values smaller than $1$. Hence, it is possible for strong lensing to produce multiple gravitational-wave images from the same source, in which some are magnified and identified as superthreshold gravitational-wave detections, and the rest being de-magnified with much weaker amplitudes that are buried within the noise background. We refer to the latter as subthreshold signals.

In this work, we are interested in searching for lensed counterparts, potentially being subthreshold, to confirmed superthreshold gravitational waves by effectively reducing the noise background in a search while conserving the targeted foreground. The following section describes the proposed method.

\section{The TESLA search method for subthreshold lensed gravitational waves}\label{Section: Methods}
In this section we will introduce the TargetEd subthreshold Lensing seArch (TESLA) pipeline aiming to search for potential subthreshold lensed counterparts to confirmed superthreshold gravitational waves. 

\subsection{The need for a reduced targeted template bank}
As explained in Section \ref{Section: Background}, a large template bank is used for a general search for gravitational waves to cover a wide parameter space, solely because we have no prior information about the parameters of the gravitational waves we are searching for. However, higher number of templates results in higher trials factors and larger noise background. This will lower the ranking statistics of gravitational wave signal, particularly those being weaker, and caused them to remain un-identified. Hence, we have to develop a way to reduce the nuisance noise background while keeping the targeted foreground constant by reducing the search parameter space, keeping only a subset of templates from the original full template bank.

\subsection{Deciding which region of the parameter space should be targeted: Signal sub-space and noise fluctuations}
The task upfront now becomes deciding the parameter space that we should be searching in to find subthreshold lensed counterparts to a given targeted superthreshold event. We argue here that there are two major contributing factors: (1) information about the signal sub-space gained from the target superthreshold event, and (2) noise fluctuations in the data, which can lead to finding a candidate event with a template whose parameters differ from those of the target event. 

\subsubsection{Information about the signal sub-space}
Recall from Section \ref{Section: Background} that strongly-lensed gravitational waves from the same source should have identical waveforms apart from (1) a relative arrival time delay, (2) an overall amplitude scaling factor, and (3) an additional Morse phase factor. That said, the potential subthreshold lensed counterparts we are searching for should have the same intrinsic parameters (e.g. component masses and spins) as the target superthreshold event. In principle, if we know precisely the underlying parameters for the target event, a single template with the exact same parameters would be ideal to search for its potential subthreshold lensed counterparts. However, the parameters of a gravitational wave are not exactly known, but instead given by best estimates from the posterior probability distribution obtained by Bayesian parameter estimation. Therefore, a good starting point would be to keep templates within the parameter space enclosed by the $90\%$ credible region of the posterior probability distribution of the target event's parameters.

Nevertheless, we argue that the posterior space is insufficient to cover all potential subthreshold lensed counterparts. Bayesian parameter estimation for gravitational waves typically assumes that noise in the data is Gaussian and stationary, which is not true in reality. That said, the posterior probability distribution obtained is for one noise realization only, i.e. the width of the posterior space does not account for noise fluctuations in actual data. Should the superthreshold signal be found at a different time in the data, the resulting posterior probability distribution obtained from Bayesian parameter estimation can be significantly different from the initial one. This argument will be demonstrated in later sections of this paper.

\subsubsection{Noise fluctuations in the data}
Should noise in actual data be stationary and Gaussian, the posterior space of the target event would be sufficient to serve as a search sub-space to look for potential subthreshold lensed counterparts. However, noise fluctuations in actual data add complexity, since they can result in the subthreshold signal being found with a template that falls outside this parameter sub-space. Hence, we will need to also consider the effects of noise fluctuations in actual data when deciding which region of the parameter space should be targeted. We do this by injecting subthreshold signals into noisy detector data and identify all the templates that can recover them; this is described in some details below.

\subsection{An injection campaign accounting for both factors}
Strongly-lensed gravitational waves from the same source should have exactly the same waveform, differing only by an overall scaling factor. Hence we can use the posterior samples obtained by Bayesian parameter estimation of the target event to generate possible simulated lensed injections that have similar parameters (i.e. component masses and spins) and sky location as the target event. To mimic the effect of lensing de-magnification, we reduce the amplitude and hence the SNRs of the injections. This can be done by increasing the source effective distance, as the optimal SNR scales inversely with the source's effective distance (see equation \ref{eqn:rho_distance}).
In detail, we take the posterior samples of the target event and rank them in decreasing order of likelihood. Within a given injection period, we generate, for each posterior sample, one injection with the original optimal SNR, and nine\footnote{The number nine is arbitrarily chosen. In principle, one can generate as many injections as one wishes.} additional weaker injections with smaller optimal SNRs by increasing their effective distances, requiring that their SNRs in each detector have to be $\geq 4$ to ensure they can be registered as a trigger during the matched-filtering process in the search. These simulated lensed injections  represent possible subthreshold lensed counterparts to the superthreshold target event. We then inject these simulated signals into actual data, and use \gstlal$\,$ to recover\footnote{An injection is considered ``recovered" if the corresponding trigger has a False-Alarm-Rate (FAR) $< 1/30$ days.} them with a general template bank. Because of noise fluctuations, some injections will be found by templates that have parameters significantly different from those within the posterior space of the target event. In the end, we keep only templates that can find these injections, and use them to construct a reduced targeted template bank to search for possible subthreshold lensed counterparts to the target event. Performing the injection campaign allows us to approximate a near-to-optimal targeted template bank taking into account both the information of the signal subspace we gained from the target event (by using the posterior samples to generate simulated lensed injections) and noise fluctuations in actual data. This ensures templates in the reduced bank can identify any potential subthreshold lensed signals while effectively reducing the noise background.

\subsection{A targeted search to dig up possible lensed candidates}
Once a targeted bank is constructed, we again use \gstlal$\,$ to search through all possible data with the targeted bank to look for potential subthreshold lensed counterparts to the target superthreshold event. As explained in Section \ref{Section: Background}, \gstlal$\,$ outputs a list of candidate events ranked by their assigned ranking statistics, including FARs and $\ln \mathcal{L}$. It is important to remind readers that the assigned FARs to the candidate events here are not measures of how likely they are to be lensed counterparts to the target event, but rather, as in the full search, we use the FAR to distinguish noise events (false alarms) from real astrophysical signals, whether or not they are lensed counterparts to a target event. In this case, we should use the ranking statistics assigned as a priority ranking for follow-up analysis to decide how likely each candidate event is a lensed counterpart to the target event. The details for the follow-up analysis are discussed in~\cite{Liu:2020par,Lo:2021nae} and are out of this paper's scope. Readers are reminded that the sole purpose of the TESLA search pipeline is to reduce the nuisance noise background effectively, and in turn up ranking possible subthreshold lensed counterparts to a target superthreshold event, assuming it is strongly lensed. It does not serve the purpose of estimating how likely the found potential candidates are indeed lensed counterparts to the target event.

\begin{figure}[h!]
\includegraphics[width=\columnwidth]{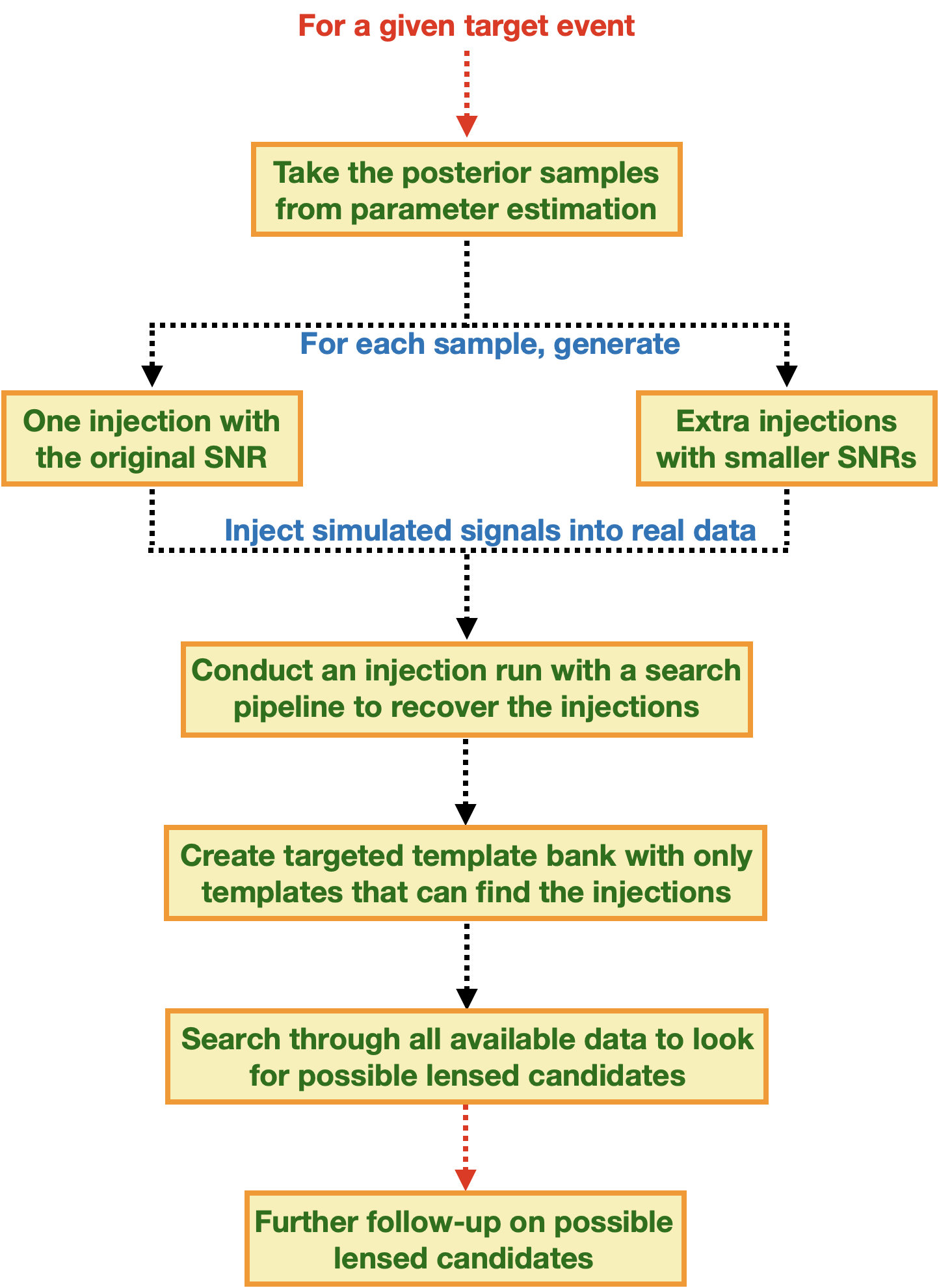}
\caption{\label{Fig: TESLA} Workflow of the targeted subthreshold search method (TESLA).}
\end{figure}

Figure~\ref{Fig: TESLA} summarizes the major steps in the TESLA search pipeline discussed in this section.

\section{Simulation campaign}\label{Section: MDC}
We perform a simulation campaign to test the effectiveness of the proposed TESLA pipeline to search for potential subthreshold lensed counterparts to a target superthreshold gravitational wave, assuming it is strongly lensed.
\begin{figure}[h!]
\includegraphics[width=\columnwidth]{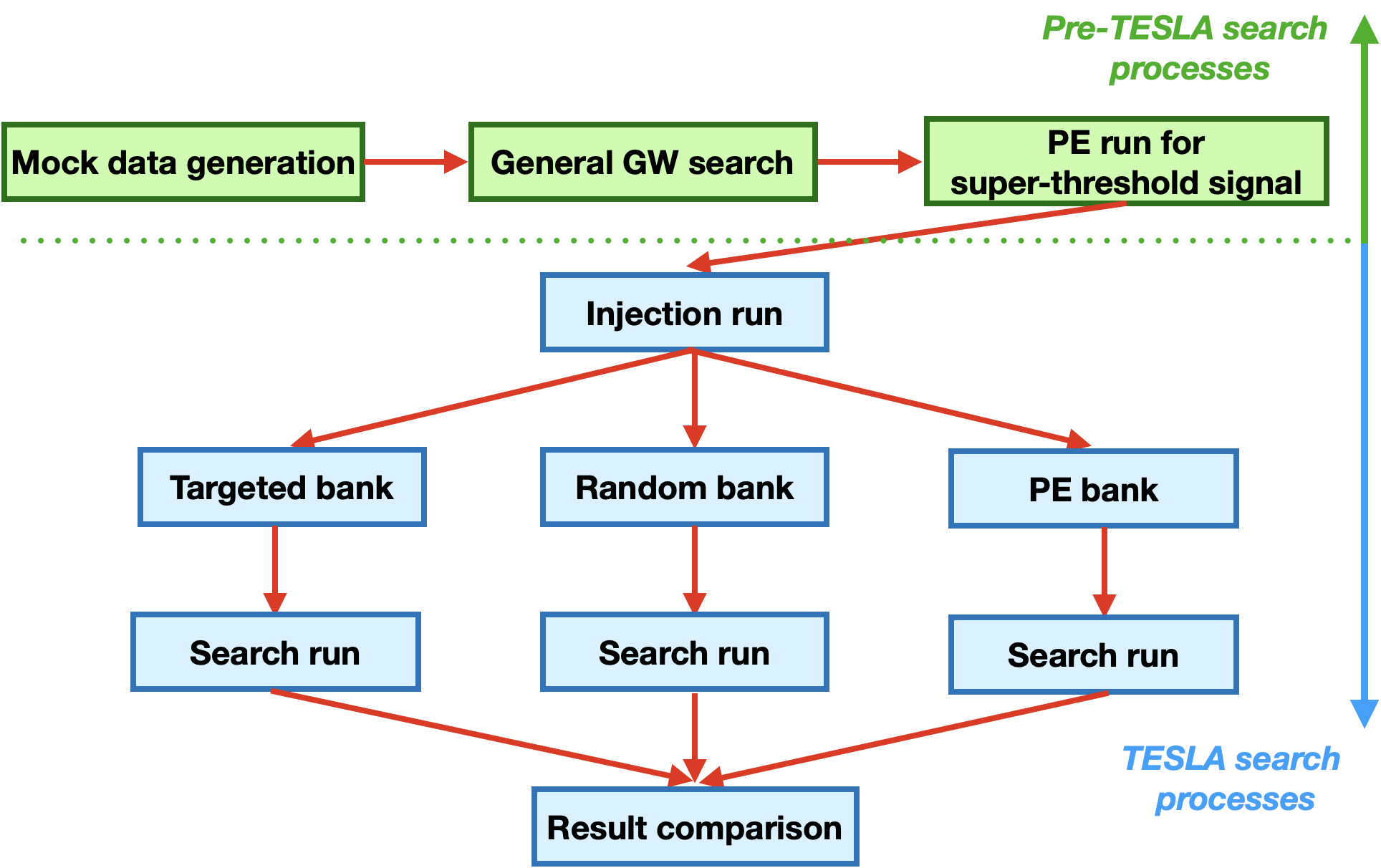}
\caption{\label{Fig: MDC analysis flow} The analysis flow of the simulation campaign.}
\end{figure}
Figure~\ref{Fig: MDC analysis flow} outlines the flow of the simulation campaign. We first prepare a mock data stream with a set of injected lensed signals, one being superthreshold and the other being subthreshold. A general search is then performed using \gstlal$\,$with the usual large general template bank\footnote{The general template bank is composed of several sub-banks targeted different systems. The minimal match of the sub-banks are in general $\geq 97\%$, with certain banks having minimal match $\geq 99\%$, see Table II in \cite{LIGOScientific:2020ibl}.}. The general search is expected to recover the superthreshold signal. Bayesian parameter estimation is then performed for the found superthreshold signal, which outputs a set of posterior samples. Then, we apply the TESLA search pipeline to perform an injection campaign and construct a targeted bank to search for the potential subthreshold lensed counterparts to the target event. Finally, we perform another search with \gstlal$\,$\footnote{A re-filtering is required only because (a) PE posteriors correspond to templates that are not in the full template bank, and (b) the results of the search with the full template bank discarded most subthreshold triggers, requiring us to re-run the search pipeline.} using the targeted bank to see if we can uprank the remaining subthreshold lensed signal that is injected. It has been suggested that extreme template banks, including
\begin{enumerate}
	\item[(1)] a single template bank with the template parameters being those of the posterior sample for the target event with the maximum posterior probability,
	\item[(2)] a PE template bank constructed by only keeping templates that lie within the $90\%$ credible region of the posterior probability distribution for the target event, and
	\item[(3)] a random template bank constructed by randomly selecting templates from the full bank,
\end{enumerate}
will have higher efficiencies than the targeted bank generated with the proposed TESLA pipeline. We therefore perform extra searches with \gstlal$\,$using the proposed banks and compare their performance.

\subsection{Mock data generation and information}
\begin{figure}[h!]
\includegraphics[width=\columnwidth]{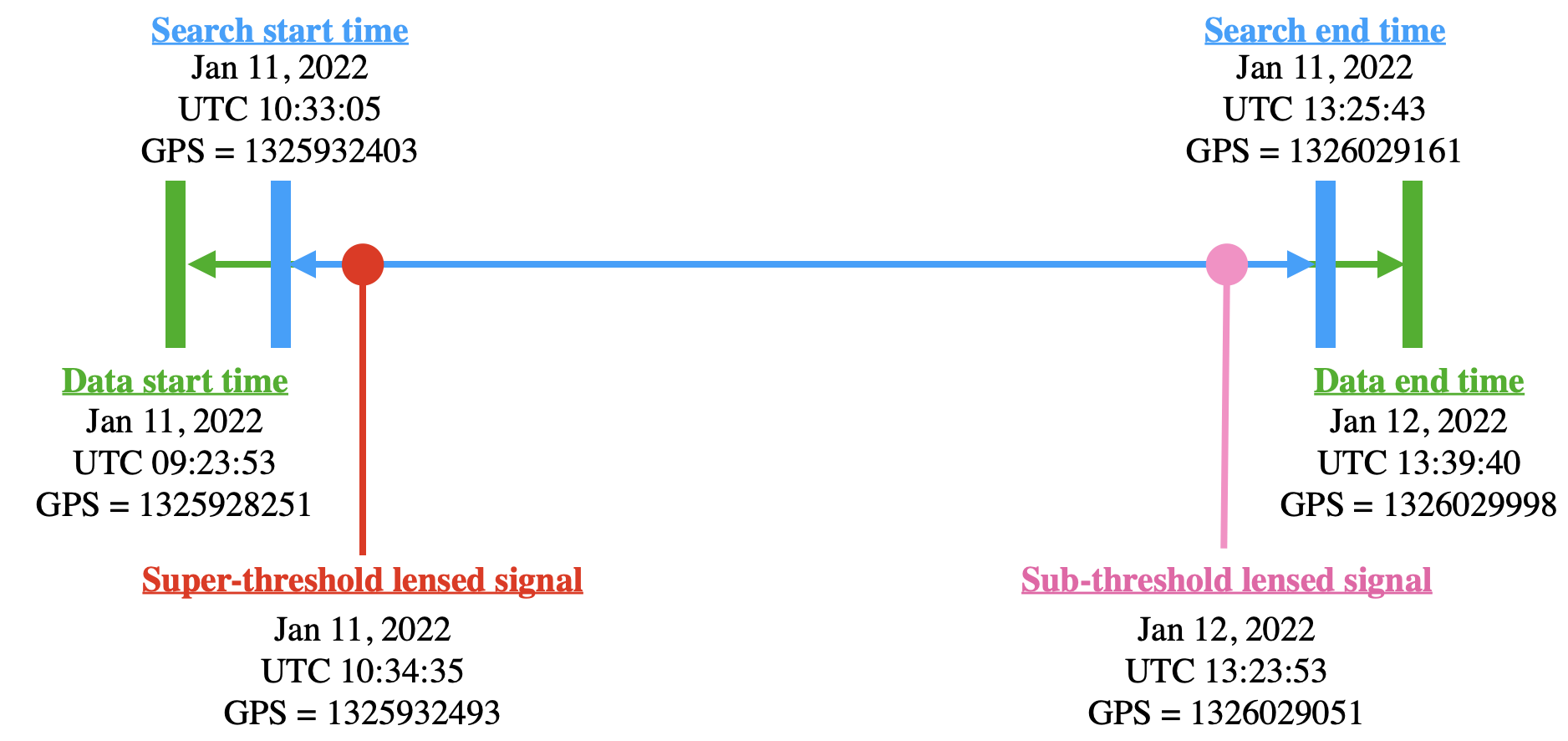}
\caption{\label{Fig: MDC data info} Information about the mock data used for the simulation campaign.}
\end{figure}
Figure~\ref{Fig: MDC data info} summarizes the basic information of the mock data stream used for this simulation campaign. For LIGO Hanford, LIGO Livingston and Virgo detector, we generate a $
\sim28$-hour-long data stream with Gaussian noise recolored with O3a characteristic power spectral densities (PSDs). We assume no detector downtime\footnote{A detector is considered ``down" if it is not in observing mode.}, and no times are vetoed. A pair of strongly-lensed gravitational waves simulated following \cite{Wierda:2021upe} is generated using the SEOBNRv4pseudoFourPN~\cite{Bohe:2016gbl} waveform approximant, and is injected into the mock data. The superthreshold signal and subthreshold signal are injected at times shown in figure \ref{Fig: MDC data info}. Details about the source parameters of the gravitational-wave signal pair are listed in table~\ref{Table: Lensed pair info}.
\begin{table}[h!]
\begin{tabular}{c c c}
\hline
\hline
Properties & Superthreshold & Subthreshold \\
& signal & signal\\
\hline
GPS time & $1325932493$ & $1326029051$\\
Magnification $\mu_i$ & $1.503$ & $-0.595$ \\
Distance (Mpc) & $2842.00$ & $4518.21$ \\
Image type & I & II\\
&&\\
Primary mass $m_1^\text{det}$ & \multicolumn{2}{c}{$42.0 M_\odot$}\\
Secondary mass $m_2^\text{det}$ & \multicolumn{2}{c}{$39.9 M_\odot$}\\
Dimensionless spins & \multicolumn{2}{c}{$\chi_{1/2,x}=\chi_{1/2,y}=0$},\\
& \multicolumn{2}{c}{$\chi_{1,z} = 0.488$, $\chi_{2,z} = -0.245$}\\
Right ascension $\alpha$ & \multicolumn{2}{c}{$2.939$} \\
Declination $\delta$ & \multicolumn{2}{c}{$0.143$} \\
Inclination $\iota$ & \multicolumn{2}{c}{$2.720$}\\
Polarization $\Psi$ & \multicolumn{2}{c}{$4.093$}\\
Source redshift $z_\text{source}$ & \multicolumn{2}{c}{0.579} \\
Lens redshift $z_\text{lens}$ & \multicolumn{2}{c}{0.245} \\
\hline
\hline
\end{tabular}
\caption{\label{Table: Lensed pair info}Information of the injected lensed gravitational-wave pair for the simulation campaign. All properties reported here are measured in the detector frame.}
\end{table}
In later parts of this paper, we may refer to the superthreshold lensed signal as MGW$220111$a and to the subthreshold signal as MGW$220112$a.

\subsection{Performing a general search}
We use \gstlal$\,$to perform a search at the times shown in figure \ref{Fig: MDC data info} following the settings used to search for gravitational waves within O3a data in GWTC-2~\cite{LIGOScientific:2020ibl}. As shown in figure~\ref{Fig: MGW220111a targeted bank}, the general template bank consists of $1412263$ templates, covering component masses between $1M_\odot$ and $400 M_\odot$, with the dimensionless spins assumed to be either aligned or anti-aligned of magnitudes $<0.999$. Template waveforms with chirp mass (detector frame) $\mathcal{M}_c^\text{det} = (m_1m_2)^{3/5}/(m_1+m_2)^{1/5} < 1.73 M_\odot$ are generated using the TaylorF2 waveform approximant~\cite{Blanchet:1995ez,PhysRevD.44.3819,Poisson:1997ha,Damour:2001bu,Mikoczi:2005dn,Blanchet:2005tk,Arun:2008kb,Buonanno:2009zt,Bohe:2013cla,Bohe:2015ana,Mishra:2016whh}\footnote{The TaylorF2 approximant only covering the inspiral is used for the BNS region because the merger and ringdown are outside of the LIGO sensitive band.}, and the rest using the SEOBNRv4\_ROM waveform approximant~\cite{Bohe:2016gbl}. As expected, the search recovers the superthreshold signal with the highest ranking statistics (FAR$=2.25\times10^{-21}$Hz, rank $1$) among all other triggers. The subthreshold signal is also registered as a trigger, but with insufficient significance (FAR$=1.53\times10^{-3}$Hz, rank $>100$) to be considered as a possible gravitational-wave signal. Table~\ref{Table: General search results} summarizes the general search results for the two injected signals.
\begin{table}[h!]
\begin{tabular}{c c c}
\hline
\hline
Search results & Superthreshold & Subthreshold \\
& signal & signal \\
\hline
GPS time & $1325932493$ & $1326029051$\\
Rank & $1$ & $>100$ \\
FAR (Hz) & $2.25\times10^{-21}$ & $1.53\times10^{-3}$ \\
$\ln \mathcal{L}$ & $43.37$ & $2.88$ \\
Network SNR $\rho_\text{network}$ & $12.1$ & $7.40$ \\
\hline
\hline
\end{tabular}
\caption{\label{Table: General search results}Results of the search for the two injected lensed signals using the general template bank.}
\end{table}
We then apply Bilby~\cite{Ashton:2018jfp,Romero-Shaw:2020owr}, a Bayesian inference library for gravitational-wave astronomy, to perform parameter estimation (PE) for the superthreshold signal, which outputs a set of posterior samples required for applying the TESLA search pipeline.

\subsection{Applying the TESLA method}
\begin{figure}[h!]
\includegraphics[width=\columnwidth]{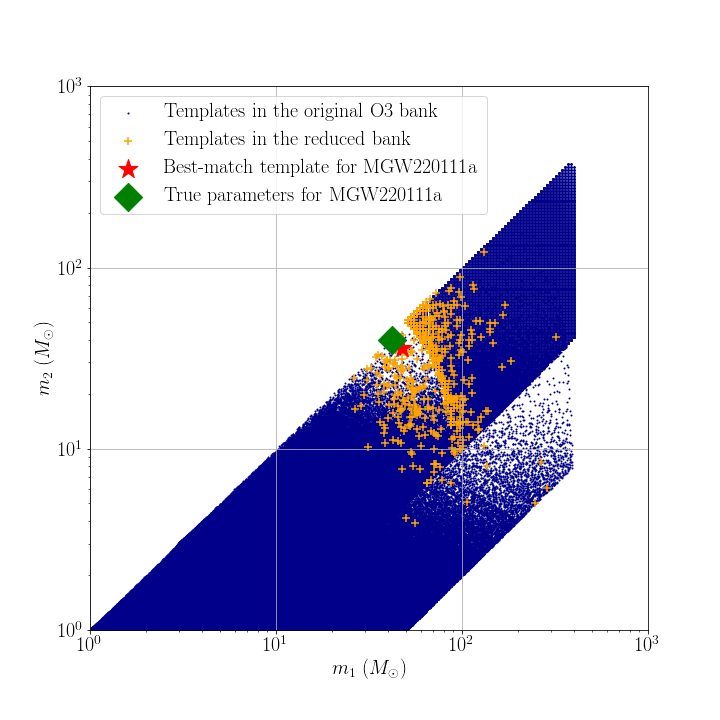}
\caption{\label{Fig: MGW220111a targeted bank} (Color online) The templates in the original and targeted bank, plotted in dark blue and orange respectively on the $m_1$-$m_2$ plane. The best-match template for MGW220111a is indicated by a red star, and the true parameters of MGW220111a is represented by a green diamond. As we can see, even when the noise in the mock data is almost stationary and Gaussian, subthreshold lensed signals can still be found by templates with parameters very different from those within the posterior space of the superthreshold target event. This demonstrates that the posterior space of the target event itself is insufficient to cover all possible subthreshold lensed counterparts.}
\end{figure}

Next, we apply the TESLA search pipeline to perform an injection campaign to construct a reduced targeted template bank to search for the remaining subthreshold lensed counterpart to the superthreshold target event. We generated $5868$ simulated lensed injections using the posterior samples obtained from the PE of the superthreshold lensed event.
We inject these simulated signals into the mock data and perform another search using \gstlal$\,$with the general template bank and try recovering them. $552$ templates are rung up by the recovered injections\footnote{Note that this does not mean only $552$ injections are recovered. The same template can be used to recover multiple injections. See Table~\ref{Table: Injection recovery} for the actual number of injections recovered.}, and they are used to construct the targeted template bank (see figure~\ref{Fig: MGW220111a targeted bank}). As we can see, even when the noise is almost stationary and Gaussian, subthreshold lensed signals can still be found by templates with parameters very different from those within the posterior space of the superthreshold target event. This demonstrates our earlier argument that the posterior space of the target event itself is insufficient to cover all possible subthreshold lensed counterparts.
Finally, we perform another search using \gstlal with the targeted template bank over the same period of mock data to try recovering the remaining injected subthreshold lensed signal. Note that we also included the lensed injection set that was used to determine the templates that we are keeping in the reduced template bank in the search for performance analysis in later sections of the paper (see section \ref{subsec: Targeted search results}). Table~\ref{Table: Targeted search results} summarizes the search results for the two injected signals using the TESLA targeted template bank.
\begin{table}[h]
\begin{tabular}{c c c}
\hline
\hline
Search results & Superthreshold & Subthreshold \\
& signal & signal\\
\hline
GPS time & $1325932493$ & $1326029051$\\
Rank & $1$ & $3$ \\
FAR (Hz) & $5.37\times10^{-21}$ & $4.27\times10^{-5}$ \\
$\ln \mathcal{L}$ & $48.63$ & $12.13$ \\
Network SNR $\rho_\text{network}$ & $12.20$ & $7.60$ \\
\hline
\hline
\end{tabular}
\caption{\label{Table: Targeted search results}Results of the targeted search of the simulation campaign for the two injected lensed signals.}
\end{table}
We can see that (1) The FAR of the subthreshold signal has been reduced by two orders of magnitude, with the log likelihood ratio $\ln \mathcal{L}$ and network SNR $\rho_\text{network}$ increased. That is, the ranking statistics of the subthreshold signal have been improved. (2) The ranking of the subthreshold signal improves significantly from its previous position of $>100$ to the current $3$. This means the TESLA search pipeline has successfully upranked the subthreshold signal, and hence made it easier to be identified as a possible gravitational wave for further analysis. 
We admit that the new FAR of the subthreshold signal still does not pass the usual FAR threshold of $1$ in $30$ days. This is primarily due to the observing time being too short. 
However, we note that the FARs assigned to each candidate here should only be treated as priority ranking for follow-up analysis to determine whether or not (1) they are gravitational waves, and (2) they are lensed counterparts of the target event. The increase in ranking of the subthreshold signal from $>100$ to $3$ demonstrates that the TESLA search pipeline is effective in reducing unwanted noise background while conserving the desired foreground, fulfilling its task to uprank potential subthreshold lensed counterparts for a targeted superthreshold event.

\subsection{Performance comparison with other suggested alternatives}
\begin{figure}[h!]
\includegraphics[width=\columnwidth]{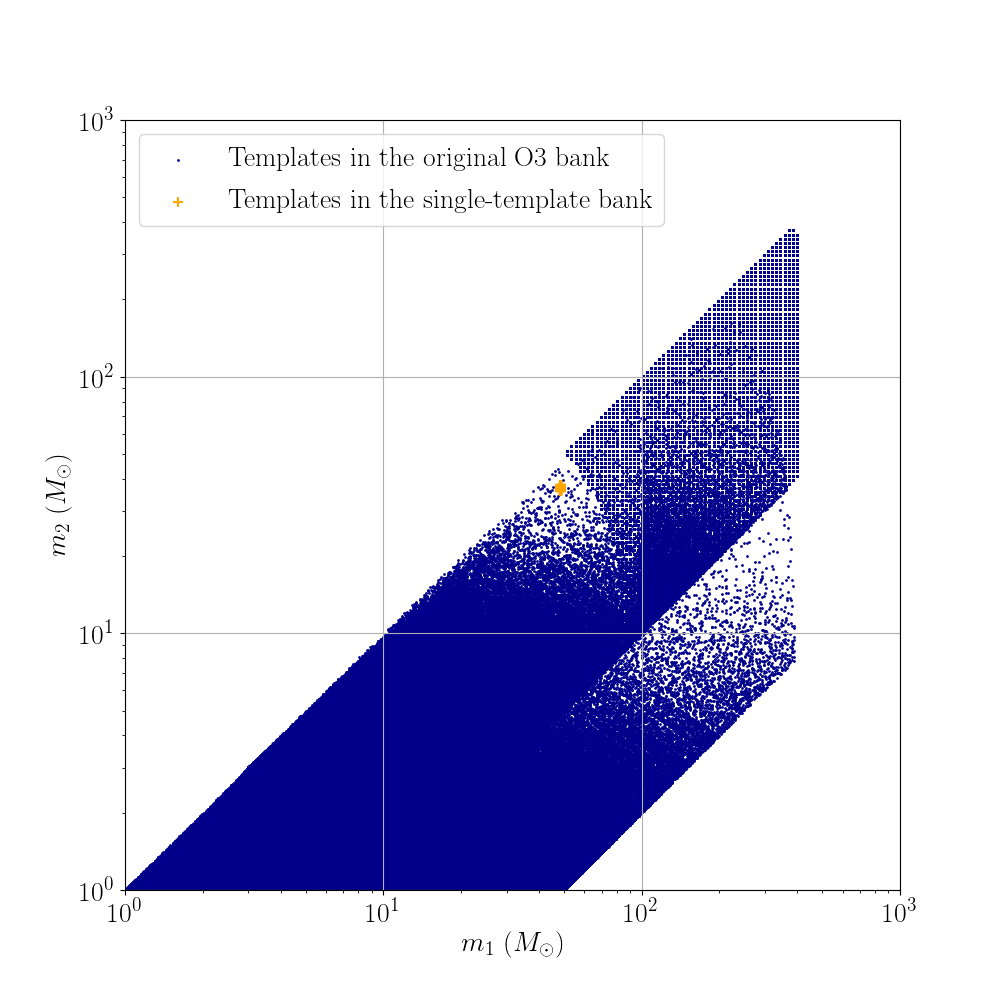}
\caption{\label{Fig: MGW220111a single-template bank} (Color online) The templates in the original and ``single-template" bank, plotted in dark blue and orange respectively on the $m_1$-$m_2$ plane. The ``single-template" bank is a bank with $100$ templates having component masses within $\pm0.1M_\odot$ from those of the posterior sample with maximum posterior to mimic the single template bank.}
\end{figure}

\begin{figure}[h!]
\includegraphics[width=\columnwidth]{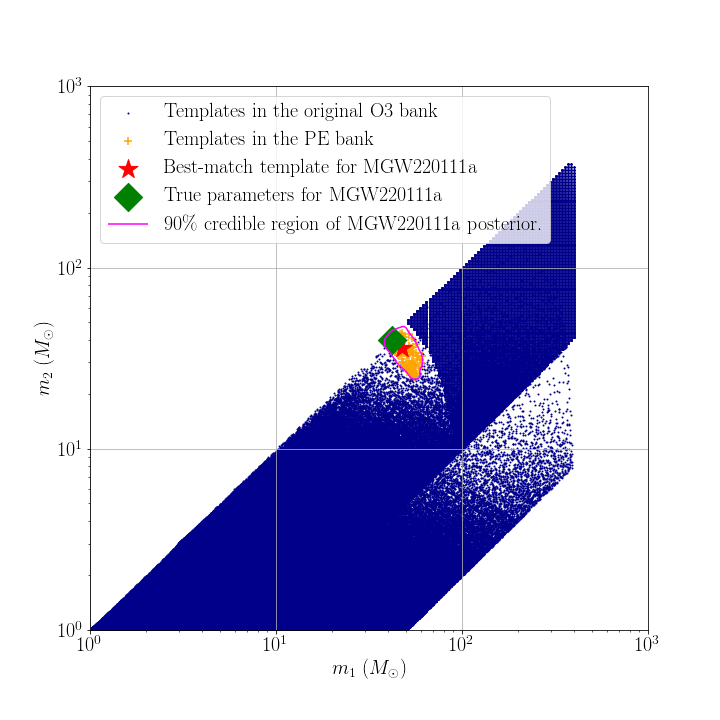}
\caption{\label{Fig: MGW220111a PE bank} (Color online) The templates in the original and PE bank, plotted in dark blue and orange respectively on the $m_1$-$m_2$ plane. The best-match template for MGW220111a is indicated by a red star, and the true parameters of MGW220111a are represented by a green diamond. The purple curve represents the boundary to the $90\%$ credible region of the posterior probability distribution for MGW220111a. The PE bank is generated by keeping only templates that lie within the $90\%$ credible region of the posterior probability distribution, containing only $81$ templates.}
\end{figure}

\begin{figure}[h!]
\includegraphics[width=\columnwidth]{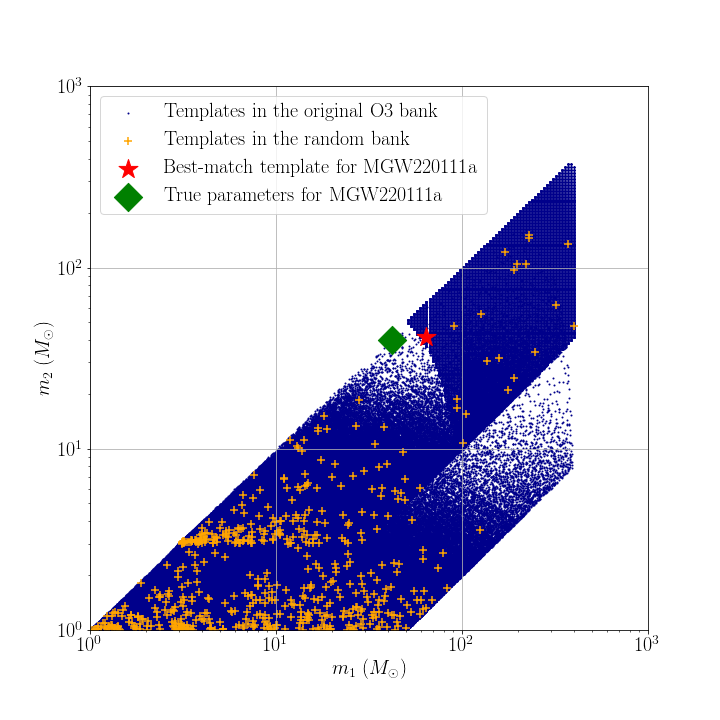}
\caption{\label{Fig: MGW220111a random bank} (Color online) The templates in the original and random bank, plotted in dark blue and orange respectively on the $m_1$-$m_2$ plane. The best-match template for MGW220111a is indicated by a red star, and the true parameters of MGW220111a are represented by a green diamond. The random bank contains the same number of templates, i.e. $552$ templates, as the targeted template bank, and they are randomly selected from the original template bank.}
\end{figure}

Suggestions have been made that (1) a single template bank, (2) a PE template bank or (3) a random template bank will be more efficient than the targeted template bank constructed with the proposed TESLA pipeline. Here we conduct additional searches using the proposed alternative banks to compare their performance. 
A random template bank is generated by randomly selecting the same number of templates (i.e. $552$ templates) as the targeted template bank. A PE template bank with $81$ templates is generated by keeping only templates that lie within the $90\%$ credible region of the posterior probability distribution obtained by Bayesian parameter estimation for the target event. A single template bank in principle should only contain one template with parameters identical to those of the posterior sample with maximum posterior of the target superthreshold event\footnote{For practical reasons, we use a bank with $100$ templates having component masses within $\pm0.1M_\odot$ from those of the posterior sample with maximum posterior to mimic the single template bank.}. Figures~\ref{Fig: MGW220111a single-template bank}, \ref{Fig: MGW220111a PE bank} and ~\ref{Fig: MGW220111a random bank} show the distribution of templates in the ``single-template" bank, PE bank and random bank respectively.

\subsubsection{Recovering the subthreshold lensed signal}
Three additional searches using \gstlal$\,$are performed over the same period of mock data as the injection run with the random template bank, the PE template bank and ``single-template" bank respectively, in order to recover the subthreshold injected signal. Tables~\ref{Table: Random search results}, \ref{Table: PE search results} and \ref{Table: Single search results} summarize the search results for the two injected signals.
\begin{table}[h!]
\begin{tabular}{c c c}
\hline
\hline
Search results & Superthreshold & Subthreshold \\
& signal & signal\\
\hline
GPS time & $1325932493$ & $1326029051$\\
Rank & $1$ & $>100$ \\
FAR (Hz) & $3.07\times10^{-14}$ & $1.54\times10^{-2}$ \\
$\ln \mathcal{L}$ & $28.0$ & $-2.14$ \\
Network SNR $\rho_\text{network}$ & $10.45$ & $7.21$ \\
\hline
\hline
\end{tabular}
\caption{\label{Table: Random search results}Results of the search for the two injected lensed signals using the random template bank.}
\end{table}

\begin{table}[h!]
\begin{tabular}{c c c}
\hline
\hline
Search results & Superthreshold & Subthreshold \\
& signal & signal\\
\hline
GPS time & $1325932493$ & $1326029051$\\
Rank & $1$ & $7$ \\
FAR (Hz) & $3.61\times10^{-5}$ & $9.05\times10^{-5}$ \\
$\ln \mathcal{L}$ & $48.13$ & $12.69$ \\
Network SNR $\rho_\text{network}$ & $12.53$ & $7.663$ \\
\hline
\hline
\end{tabular}
\caption{\label{Table: PE search results}Results of the search for the two injected lensed signals using the PE template bank.}
\end{table}

\begin{table}[h!]
\begin{tabular}{c c c}
\hline
\hline
Search results & Superthreshold & Subthreshold \\
& signal & signal\\
\hline
GPS time & $1325932493$ & $...$\\
Rank & $1$ & $...$ \\
FAR (Hz) & $6.11\times10^{-6}$ & $...$ \\
$\ln \mathcal{L}$ & $24.11$ & $...$ \\
Network SNR $\rho_\text{network}$ & $12.49$ & $...$ \\
\hline
\hline
\end{tabular}
\caption{\label{Table: Single search results}Results of the search for the two injected lensed signals using the ``single-template" bank.}
\end{table}
From the results, we see that (1) The ``single-template" bank fails to even register the subthreshold signal as a trigger in the first place during the matched-filtering process, (2) the PE template bank successfully upranks the subthreshold signal to a rank $7$ candidate and improves its ranking statistics, but its performance is not as good as compared to that using the TESLA search pipeline, and (3) the random template bank fails to improve the ranking and the ranking statistics of the subthreshold event. This means that the targeted foreground is affected by the reduction in the number of templates for the random template bank. It is therefore evident that the random bank is not suitable to search for potential subthreshold lensed gravitational waves. 

\subsubsection{Simulated lensed injections recovery}\label{subsec: Targeted search results}
To further compare the performance of the banks, we analyse the change in number of lensed injections recovered\footnote{As before, an injection is considered ``recovered" if the corresponding trigger has a FAR $< 1/30$ days.} using the four proposed banks as compared to using the general template bank. Table~\ref{Table: Injection recovery} summarizes the findings. 
\begin{table}[h]
\begin{tabular}{c c c c c c}
\hline
\hline
Injections & General & TESLA & Random & PE & Single \\
\hline
Total & $5868$ & $5868$ & $5868$ & $5868$ & $5868$ \\
Found & $1793$ & $1959$ &  $402$ & $299$ & $1076$ \\
Missed & $4075$ & $3909$ & $5466$ & $5569$ & $4792$\\
Found $\%$ change & - & $+9.26\%$ & $-77.5\%$ & $-80.3\%$ & $-40.0\%$\\
\hline
\hline
\end{tabular}
\caption{\label{Table: Injection recovery}Number of injections found and missed during the search of mock data using the general template bank, TESLA targeted template bank, random template bank, PE template bank and ``single-template" bank respectively.}
\end{table}
The targeted template bank constructed using the TESLA search pipeline succeeds in recovering more lensed subthreshold injections than the other banks. The ``single-template" bank, the PE template bank and the random template bank miss more lensed subthreshold injections. The random template bank is expected not to give a satisfactory performance in recovering the lensed subthreshold injections. While the lensed subthreshold injections are generated using the exact same parameters as the posterior samples of the target event (i.e. the injections should all have similar parameters as to the templates in the PE template bank), the PE template bank misses even more injections as compared to using the general template bank\footnote{In fact, it misses even more injections than the random bank, but this should not be alarming. Given that the injections are subthreshold, they are more likely to be recovered by templates with very different parameters than their true parameters. The random bank, while being completely random, covers a much wider parameter space than the PE bank, and hence have a higher chance in recovering the subthreshold injections.}. This again demonstrates our argument that constructing the targeted template bank solely by considering the posterior signal sub-space of the superthreshold target event is insufficient. In this simulation campaign, we are simply fortunate that the PE bank can recover the injected subthreshold lensed signal. Should the injected subthreshold signal be even weaker, or should it be injected at a time at which noise is very different from that around the superthreshold signal, the PE template bank is more likely to miss it. On the other hand, the targeted template bank created using the TESLA pipeline is more likely to recover it since the bank is constructed by considering both information about the signal sub-space gained from the target event as well as noise fluctuations in the data.

\subsubsection{Sensitive range at different FAR threshold}
\begin{figure}[h]
\includegraphics[width=\columnwidth]{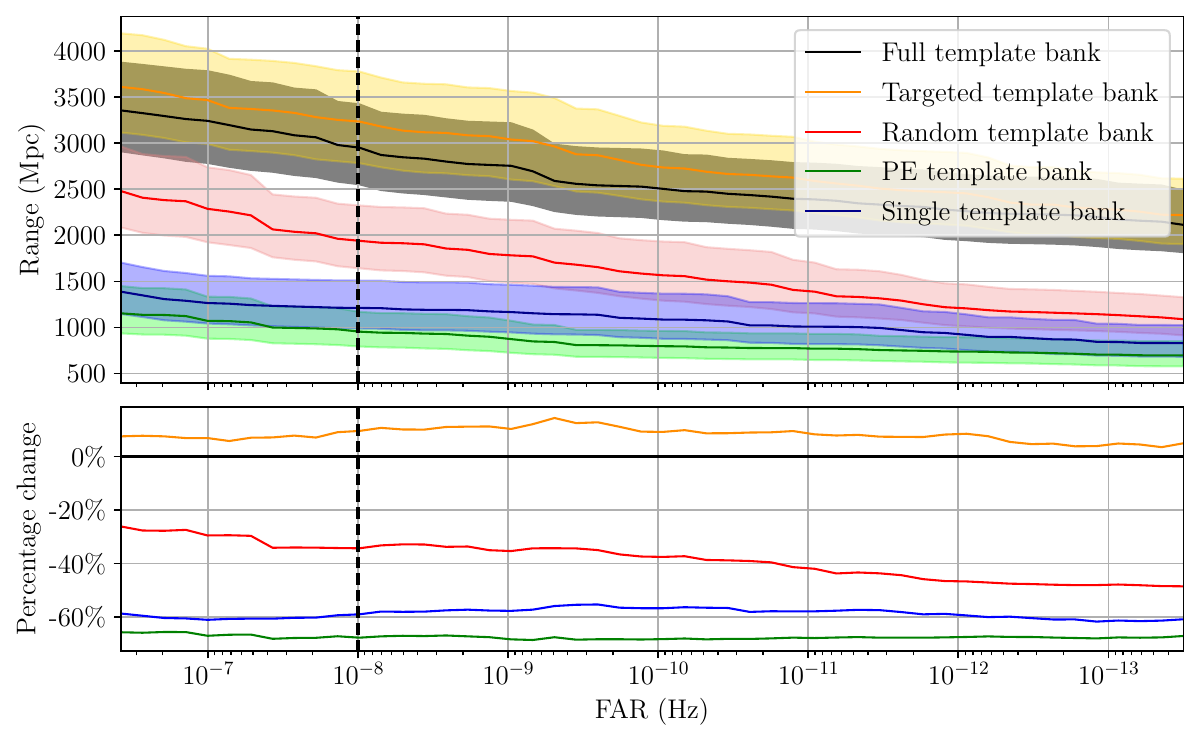}
\caption{\label{Fig: Range vs FAR} (Color online) (Top panel) The sensitive range v.s.~FAR for MGW220111a-alike signals using the full template bank (black), targeted template bank (orange), ``single-template" bank (blue), PE template bank (green) and random template bank (red) respectively. The shaded band for each curves represent the corresponding $1$-sigma region. (Bottom panel) The corresponding percentage changes in sensitive range v.s.~FAR for the different banks. The percentage-change curve (orange) representing results using the targeted bank constructed by the TESLA search pipeline is above that of the full template bank, showing improvement in terms of sensitivity towards MGW220111a-alike (lensed) subthreshold signals. The vertical striped line shows the typical FAR threshold for triggers below which we consider as possible lensed candidates.}
\end{figure}

Finally, we use the $5868$ lensed subthreshold injections\footnote{These are the same injections used in the simulation campaign to create the reduced template bank.} to evaluate the sensitive range\footnote{The sensitive range is the distance out to which we may identify gravitational waves averaged over relevant parameters including sky location and binary orientation. Note that in this analysis we assumed the injections are not lensed, i.e. they have magnification $\mu=1$.} at different combined FAR threshold for each template bank. Figure~\ref{Fig: Range vs FAR} shows the percentage changes in sensitive range v.s.~FAR curves obtained using the alternative banks as compared to that using the full template bank for lensed subthreshold signals that are similar to the target superthreshold event.  
We can see that the percentage-change curve representing results using the targeted bank constructed by the TESLA search pipeline is above that of the full template bank, showing improvement in terms of sensitivity towards MGW220111a-alike (lensed) subthreshold signals. Meanwhile, the same curves for the random bank, the PE bank and the ``single-template bank" are below that of the full template bank, showing that the sensitivity towards MGW220111a-alike (lensed) signals is worsened. This further demonstrates that the targeted template bank created using the TESLA search pipeline has the best performance among the four banks to search for potential subthreshold lensed gravitational waves for a target superthreshold event.

\subsubsection{Summary of results}
To sum up the results presented above, in this simulation campaign we investigated four proposed banks to search for possible subthreshold lensed counterparts of a given superthreshold gravitational wave. Three of the cases are found to be inferior in performance compared with the TESLA bank, namely the single-template bank, the PE template bank and the random template bank. The results show that none of the three alternative cases can outperform an intermediate template bank created based on the TESLA pipeline in terms of search sensitivity and effectiveness.

It should be noted that in this simulation campaign, we considered the case where lensing creates a pair of repeated gravitational-wave signals from the same source, separated by roughly a day ($~1.11$ days). In practice, the relative time delay between repeated signals can range from minutes to months for galaxy lenses. We will need to perform the injection campaign over a longer time range with a larger number of injections. This will result in an increase in size for the reduced template bank, and may affect the performance of the reduced template bank. The ranking statistics, in particular the FAR of the triggers, will also be affected based on the number of templates we have in the reduced template bank. Future work will investigate how to fine tune the selection procedure for templates included in the targeted template bank in order to find the optimal balance between coverage and sensitivity.

\section{Concluding Remarks}\label{Section: Conclusion}
The LVK collaboration has recently published the first full-scale analysis to search for lensing signatures of gravitational waves within the first half of LIGO/Virgo third observing run O3a~\cite{LIGOScientific:2021izm}, and concluded that there is not yet any compelling evidence for gravitational lensing of gravitational waves. One featured analysis in the paper explores the possibility of strong lensing producing magnified superthreshold gravitational-wave signals, and de-magnified subthreshold copies that have insufficient significance and remain un-identified as detections. Two independent search methods were applied to search for the latter potential subthreshold lensed signals, one being the \gstlal-based TargetEd subthreshold Lensing seArch (TESLA) pipeline. 

In a general search for gravitational waves, a large template bank covering a wide parameter space is used since we have no prior information regarding the parameters of gravitational waves we are searching for. The large number of templates used contributes a high trials factors. This may bury potential subthreshold (lensed) gravitational waves in the huge noise background. To search for possible subthreshold lensed counterparts to superthreshold confirmed gravitational waves, we need to reduce the noise background while keeping the targeted foreground constant. In other words, we want to lower the noise background by tactically focusing a particular region in the parameter subspace, while keeping the targeted foreground constant, and hence upranking any potential subthreshold lensed candidates to the superthreshold target events.

In this paper we explain the methodology of the TESLA pipeline in detail, and demonstrate that the TESLA pipeline can efficiently search for possible subthreshold lensed counterparts to confirmed superthreshold gravitational-wave detections.

The TESLA pipeline fulfils the task by conducting an injection campaign. It prepares simulated lensed injections based on posterior samples obtained from Bayesian parameter estimation of the superthreshold target event, such that they have similar intrinsic parameters as the target event, but with varying effective distances and hence weaker amplitudes to mimic the de-magnifying effect caused by gravitational lensing. These injections are then injected into actual data\footnote{Note that the results in this paper make use of Gaussian simulated data, but for the actual subthreshold search, these injections are made into actual data.} and a \gstlal$\,$search is performed using the general template bank to recover these injections. Templates that can find the injections are used to construct a targeted template bank, which is then used to perform another \gstlal$\,$search to look for possible subthreshold lensed counterparts to the target event, should it be strongly lensed. We argue that the TESLA search pipeline can generate a template bank that performs better than alternatives to search for these potential subthreshold signals, as it accounts for both information about the signal-subspace gained from the target event, as well as noise fluctuations in actual data.

To assess the performance of the TESLA search pipeline, we conducted a simulation campaign in which we simulated LHO, LLO, and Virgo data streams with Gaussian noise recolored with O3a representative power spectral densities (PSDs) and a pair of lensed signals, one being superthreshold and the other being subthreshold. We first perform a \gstlal$\,$ search using the general template bank to recover the superthreshold signal, and perform Bayesian parameter estimation to generate a set of posterior samples. Then, we use the TESLA pipeline and try to recover the remaining subthreshold lensed signal from the mock data. Our results show that the TESLA pipeline can effectively uprank the subthreshold signal, improving the probability that it will be identified as a gravitational wave, and with further analysis, a lensed counterpart to the target superthreshold event.

We also compare the performance of the targeted template bank constructed with the TESLA search pipeline to suggested alternative template banks: (1) a single template bank with the template parameters being those of the posterior sample for the target event with the maximum posterior probability, (2) a PE template bank constructed by keeping only templates from the general bank that lie within the $90\%$ credible region of the posterior space for the target event, and (3) a random template bank constructed by randomly selecting templates from the general template bank. We show, by considering their performance in (1) recovering the injected subthreshold lensed signal, (2) recovering the simulated lensed injections and (3) their sensitive range for gravitational waves that are similar to the target events, that the targeted template bank constructed using the TESLA search pipeline outperforms the other three alternative banks. In fact, the results show that one would not expect additional improvement when further narrowing the template bank.

The search sensitivity of the TESLA search pipeline can be further improved. For instance, since we are looking for lensed counterparts of targeted events, using the target's sky location, we should be able to set a consistent range for the difference in arrival time and phase between participating detectors for the lensed counterparts. This will be discussed in a future paper under development. Also, the selection procedure for templates included in the targeted template bank may require further tuning to find the optimal balance between coverage and sensitivity.

This method is intended to be applied to subsequent searches for subthreshold lensed events, in future LVK papers.

\begin{acknowledgements}
The authors acknowledge the generous support from the National Science Foundation in the United States. The authors would also like to acknowledge Jonah Kanner and Bruce Allen for their useful suggestions. R. K. L. L. and T. G. F. L. would also like to gratefully acknowledge the support from the Croucher Foundation in Hong Kong. The work described in this paper was partially supported by a grant from the Research Grants Council of the Hong Kong (Project No. CUHK 14306218) and the Direct Grant for Research from the Research Committee of the Chinese University of Hong Kong. S. S. was supported in part by the LIGO Laboratory and in part by the Eberly Research Funds of Penn State, The Pennsylvania State University, University Park, Pennsylvania 16802, USA. J.C.L.C. acknowledges support from the Villum Investigator pro- gram supported by VILLUM FONDEN (Grant No. 37766) and the DNRF Chair, by the Danish Research Foundation. The authors are also grateful for computational resources provided by the LIGO Laboratory and supported by National Science Foundation Grants No. PHY-0757058 and No. PHY-0823459. This research has made use of data, software and/or web tools obtained from the Gravitational Wave Open Science Center (https://www.gw-openscience.org) \cite{2015JPhCS.610a2021V}, a service of LIGO Laboratory, the LIGO Scientific Collaboration and the Virgo Collaboration. LIGO was constructed by the California Institute of Technology and Massachusetts Institute of Technology with funding from the National Science Foundation and operates under cooperative agreement PHY-0757058.
Virgo is funded by the French Centre National de Recherche Scientifique (CNRS), the Italian Istituto Nazionale della Fisica Nucleare (INFN) and the Dutch Nikhef, with contributions by Polish and Hungarian institutes.
This paper carries LIGO Document Number LIGO-\DocumentID{}. A.K.Y.L. and R.K.L.L. would like to gratefully acknowledge the support from the National Science Foundation through the Grants NSF PHY-1912594 and NSF PHY-2207758.
\end{acknowledgements}
\bibliographystyle{apsrev4-1}
\bibliography{citations}

\end{document}